\documentclass[10pt, conference]{IEEEtran}

\usepackage{mypreamble}
\begin{document}

\title{Code Readability in the Age of Large Language Models: An Industrial Case Study from Atlassian}

\makeatletter
\newcommand{\newlineauthors}{%
  \end{@IEEEauthorhalign}\hfill\mbox{}\par
  \mbox{}\hfill\begin{@IEEEauthorhalign}
}
\makeatother

\author{
\IEEEauthorblockN{Wannita Takerngsaksiri\IEEEauthorrefmark{1}, Chakkrit Tantithamthavorn\IEEEauthorrefmark{1}, Micheal Fu\IEEEauthorrefmark{2}, Jirat Pasuksmit\IEEEauthorrefmark{3}, Kun Chen\IEEEauthorrefmark{4}, Ming Wu\IEEEauthorrefmark{4}}
\IEEEauthorblockA{
\IEEEauthorrefmark{1}Monash University, Australia. 
}
\IEEEauthorblockA{
\IEEEauthorrefmark{2}The University of Melbourne, Australia. 
\IEEEauthorrefmark{3}Atlassian, Australia. 
\IEEEauthorrefmark{4}Atlassian, United States.
\\
Contact: chakkrit@monash.edu
}
}

% \and
% \IEEEauthorblockN{Micheal Fu}
% \IEEEauthorblockA{
% The University of Melbourne, Australia}
% \and
% \IEEEauthorblockN{Chakkrit Tantithamthavorn}
% \IEEEauthorblockA{
% Monash University, Australia}
% \and
% \newlineauthors
% \IEEEauthorblockN{Jirat Pasuksmit}
% \IEEEauthorblockA{
% Atlassian, Australia}
% \and
% \IEEEauthorblockN{Kun Chen}
% \IEEEauthorblockA{
% Atlassian, United States}
% \and
% \IEEEauthorblockN{Ming Wu}
% \IEEEauthorblockA{
% Atlassian, United States}

\maketitle

\begin{abstract}
    Software engineers spend a significant amount of time reading code during the software development process, especially in the age of large language models (LLMs) that can automatically generate code.
    % This trend is amplified by the emergence of large language models (LLMs) that automatically generate code.
    % increasingly shifting the programmer's role from creator to overseer.
    However, little is known about the readability of the LLM-generated code and whether it is still important from practitioners’ perspectives in this new era.
    In this paper, we conduct a survey to explore the practitioners’ perspectives on code readability in the age of LLMs and investigate the readability of our LLM-based software development agents framework, \emph{HULA}, by comparing its generated code with human-written code in real-world scenarios.
    Overall, the findings underscore that (1) readability remains a critical aspect of software development; (2) the readability of our LLM-generated code is comparable to human-written code, fostering the establishment of appropriate trust and driving the broad adoption of our LLM-powered software development platform.

\end{abstract}

\section{Introduction}\label{sec_introduction}

% For example, Buse and Weimer~\cite{buse2008metric} defined \textit{code readability} as \textit{``a human judgment of how easy a text is to understand"}.

Code readability has long been a key focus in software engineering research~\cite{tashtoush2013impact, coleman2018aesthetics, fakhoury2018effect, mannan2018towards, alawad2019empirical, johnson2019empirical}. 
For example, Minelli~\ea~found that programmers dedicate approximately 70\% of their time to reading and understanding source code~\cite{minelli2015know}. 
Similarly, as Guido Van Rossum, the creator of Python, has noted: \emph{``Code is read more often than it is written''}\cite{pepeight2001}. 
Thus, writing readable code not only facilitates better comprehension of its functionality~\cite{johnson2019empirical} but also significantly reduces software maintenance costs.

In the age of large language models (LLMs) for code, these models have demonstrated remarkable results across various software development tasks~\cite{zhao2023survey} (e.g.,  code completion~\cite{takerngsaksiri2024syntax, takerngsaksiri2024student}, test case generation~\cite{ takerngsaksiri2024tdd,alagarsamy2024a3test}, code review automation~\cite{li2022codereviewer, pornprasit2024gpt,thongtanunam2022autotransform,hong2022commentfinder}, vulnerability detection and repair~\cite{fu2022linevul,fu2024aibughunter,fu2022vulrepair}).
Recently, there has been a growing interest in integrating LLMs into modern Integrated Development Environments (IDEs) (e.g., GitHub Copilot~\cite{copilot2024}) to assist software engineers in implementing new features, updating patches, improving code quality, fixing bugs, and resolving software development tasks.
% Since code is now being generated by Large Language Models (LLMs)\cite{adam2004}.

At Atlassian, code readability has always been a cornerstone of software development, as it plays a critical role in facilitating software evolution and maintenance. 
In the age of large language models (LLMs), like those integrated into Jira (a task management platform)~\cite{jiraai2024}, the importance of code readability is amplified. 
Readable code ensures that teams can efficiently collaborate, debug, and enhance their software, reducing technical debt and long-term costs.
Recently, we introduced HULA, a human-in-the-loop software development agents framework powered by large language models~\cite{takerngsaksiri2024human}. 
While HULA aims to streamline development tasks, a crucial question arises: does readability retain its value in this LLM-assisted paradigm, and if so, why? Additionally, how readable is the code it generates compared to human-written code? 
Understanding and ensuring code readability in LLM-generated code is vital for Atlassian, since it directly impacts the productivity and satisfaction of software development teams, enabling teams to work efficiently.

\textit{In this paper}, we aim to investigate the practitioners' perceptions on the importance, the challenges, and the state of practice of code readability, and investigate the readability of LLM-generated code and human-written code in the context of enterprise software development tasks.
Thus, we address the following two RQs.

% Through a user study of 118 participants and an investigation of the readability of HULA-generated code 
% First, we conduct an online survey with 118 participants to study the practitioners' perceptions of code readability in four main topics: the importance, the challenges, the state of practice, and the readability of LLM-generated code.
% First, we revisit the factors influencing the importance and challenges of code readability improvement in professional practices.
% Then, we explore how practitioners improve code readability in their professional practices, and how they perceive LLM-generated code compared to human-written code.
% Additionally, to deepen our understanding of the readability of LLM-generated code, we show a case study comparing the code readability of our LLM-based agent framework, \emph{HULA}, and humans.
% The experiment is run on HULA with a GPT-4 based version, using our internal datasets consisting of six widely used programming languages (i.e., TypeScript, Java, Kotlin, Python, Go and Scala).
% We address the following research questions.

\begin{figure*}[t]
    \centering
    \includegraphics[width=\textwidth]{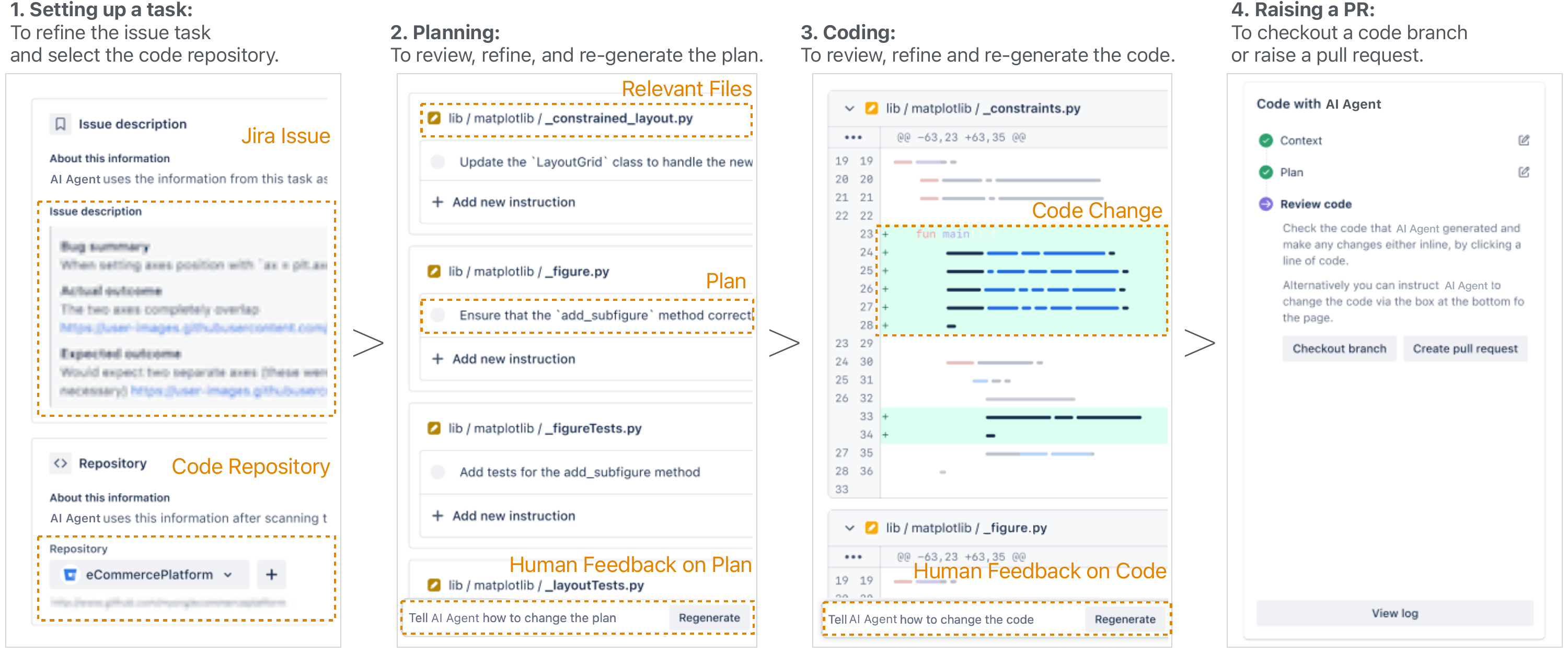}
    \caption{The user interface of our HULA (\textbf{Hu}man-in-the-\textbf{L}oop Software Development \textbf{A}gent) framework, seamlessly integrated into Atlassian JIRA~\cite{takerngsaksiri2024human}.}
    \label{fig:autodev-workflow}
\end{figure*}

\begin{enumerate}[label=\textbf{RQ\arabic*)}, left=0pt]
    \item \textbf{\rqone} \newline
    Through an online survey of 118 practitioners, we found that 81\% of practitioners agreed that code readability is important.
    The key motivation is to reduce maintenance costs in the long term, while the key challenging factor is time constraints.
    Although code readability is currently improved via code review comments, 72\% of practitioners agreed to consider adopting LLMs as an alternative.
    39\% of practitioners perceived that LLM-generated code is more readable than human-written code in general, followed by 34\% perceived that the readability of both LLM-generated and human-written code is similar.
    These findings suggest that code readability is still critically important in the age of large language models. 
    % although code can be automatically generated in the age of large language models, code readability is still critically important.
    % 56\% of practitioners answered spending more than half of their time reading code. 
    % While the top motivation to prioritize code readability is to reduce maintenance costs, the top challenging factor is the time constraint of the project.
    % In addition, although the peer review comments are still a dominant method for code readability improvement with 41\% selection, 72\% of practitioners agree to consider adopting Generative AI as an option.
    % Lastly, by showing a code pair of anonymous origin, 59\% of practitioners perceived that the ChatGPT code is more readable than the human-written code.
    % The results indicate that the readability of generative AI-produced code tends to be perceived positively, with practitioners recognizing its readability as comparable to, if not better than, human-written code.
    
    \item \textbf{\rqtwo} \newline
    Through an examination of the readability of code generated by our LLM-powered code generation (HULA~\cite{takerngsaksiri2024human}, powered by GPT-4~\cite{openai2023gpt4}) and human-written code using our 144 internal Jira software development tasks, we found that the readability of LLM-generated code is comparable to human-written code (with a negligible to small effect size difference) based on various measurements (e.g., lines of code, code complexity, comment to code ratio).
    These findings demonstrate that the LLM-generated code can closely resemble human-written code in terms of readability.
    % Specifically, LLM-generated code closely resembles human-written code across all code readability measurements with only minor differences in effect size.
    % These results indicate the promising performance of LLM in terms of code readability.
    % On the larger scale of our empirical study, the results highlight that Generative AI typically generates longer code, with more lines of code and comments, and a lower maintainability index compared to human-written code.
    % This effect is more pronounced in open-source datasets, while the proprietary dataset shows only a small effect size.
    % On the other hand, we observe no statistically significant difference in code complexity between the Generative AI-produced code and human-written code. 
    % The results indicate that the readability of the Generative AI code is influenced more by the code length than by intrinsic code complexity.

\end{enumerate}

\textbf{Contributions.} The paper contributions are as follows:
\begin{itemize}
    \item We conduct the first survey capturing practitioners' perceptions of the current state of code readability and code readability of LLM-generated code, revealing generally positive feedback on code readability when adopting HULA in software development.
    \item We revisit the importance and factors related to code readability from literature through the survey, aiming to raise awareness of code readability in practice.
    \item We conduct the first empirical study of code readability comparison between human-written code and LLM-generated code in a real-world setting using our HULA framework, revealing promising results of the LLMs.
\end{itemize}

The results lead us to conclude that the readability of LLM-generated code is comparable to that of human-written code.
While the survey reveals that the majority of practitioners perceive LLM-generated code as more readable or equally readable as human-written code, the empirical study supports the statements showing comparable results ranging from small to negligible effect sizes across multiple evaluation metrics.

To support the open science initiatives and increase the verifiability of our study, the survey and run results are available at \url{https://github.com/awsm-research/CodeReadability-GenAI}.

\textbf{Paper Organization.} 
Section~\ref{sec_background} describes the background of HULA, our LLM-based software development agents framework.
Section~\ref{sec_related_works} presents the motivation and related work. 
Section~\ref{sec_exp_design} and Section~\ref{sec_exp_result} present the approach and results of RQ1 and RQ2, respectively.
Section~\ref{sec:threat} discloses the threats to the validity of our study, while Section~\ref{sec_conclusion} draws the conclusion.

% In fact, the code readability does not depend on the origin of the code (e.g., produced by Generative AI or written by humans).

% relies on multiple factors, such as code complexity, code length 

% Finally, although we are not able to disclose all data of the experiment due to industrial confidentiality reasons, we would like to support the open science initiatives and increase the verifiability of our study, 

% The results lead us to conclude that the readability of Generative AI-produced code can be comparable to that of human-written code.
% While the empirical study indicates that Generative AI tends to generate longer code, its complexity remains similar to human-written code. 
% Notably, the survey reveals that most practitioners perceive Generative AI-produced code as more readable than human-written code. 
% Although code verbosity might be a potential concern according to the empirical study, we observe little negative feedback on code readability from the practitioners' perspective.
\section{Background}\label{sec_background}

\begin{figure}
    \centering
    \includegraphics[width=\linewidth]{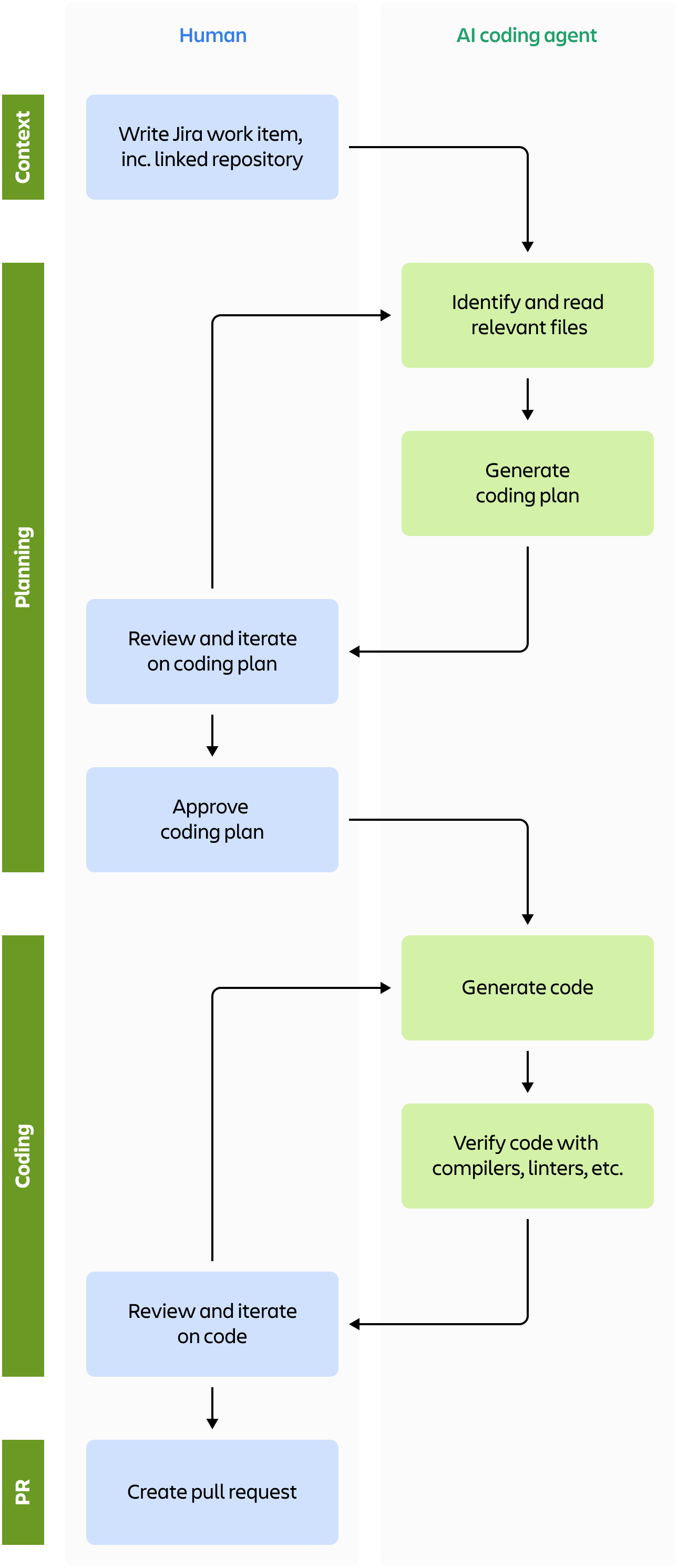}
    \caption{An Overview of HULA (\textbf{Hu}man-in-the-\textbf{L}oop Software Development \textbf{A}gents)~\cite{human2025atlassian}, seamlessly integrated into Atlassian JIRA Software.}
    % \caption[Caption for LOF]{The Overview of our HULA (\textbf{Hu}man-in-the-\textbf{L}oop Software Development \textbf{A}gent) framework, seamlessly integrated into Atlassian JIRA.\protect\footnotemark}
    \label{fig:atlassian-hula}
\end{figure}

In this section, we briefly describe our Human-in-the-loop LLM-based software development agents framework (HULA)~\cite{takerngsaksiri2024human}.
It is worth noting that in this study, all the human feedback and guidance are discarded to promote the fair comparison between human-written code and HULA-generated code.
Figure~\ref{fig:autodev-workflow} shows the user interface of HULA seamlessly embedded in Atlassian JIRA Software.

\subsection{\textbf{HULA}: \textbf{Hu}man-in-the-\textbf{L}oop Software Development \textbf{A}gents}

At Atlassian, we prioritize human expertise to guide and govern AI agents.
Rather than pursuing full automation of software development tasks, we designed an LLM-based software development agent to work alongside practitioners as a collaborative assistant.
This approach fosters Human-AI synergy~\cite{lo2023synergy,Hoda2023augmented}, where practitioners and AI agents jointly contribute to task completion, ensuring a collaborative software engineering workflow.
% To achieve this vision, we seamlessly integrate our HULA framework into Atlassian JIRA as shown in Figure~\ref{fig:autodev-workflow}, allowing practitioners to provide feedback and direction throughout the process. 

Below, we present the overview process of our HULA framework shown in Figure~\ref{fig:atlassian-hula}.
More information can be found in our official announcement\footnote{https://www.atlassian.com/blog/atlassian-engineering/hula-blog-autodev-paper-human-in-the-loop-software-development-agents}\cite{human2025atlassian} and previous publication~\cite{takerngsaksiri2024human}.

\subsubsection{Set Context}
The workflow starts when a software engineer selects a development task (i.e., a JIRA issue) along with an associated code repository. 
The description of a JIRA issue typically includes key information to help humans understand the task to be done, such as user stories, definitions of done, acceptance criteria, or example code.

\subsubsection{Planning}
The AI coding agent starts to identify the relevant files in the given repository and generates a coding plan that outlines the necessary modifications.
These include a list of relevant files and the specific changes required to address the work item. 
The software engineer reviews this plan and may either edit the generated plan directly or provide feedback to the AI coding agent for refinement and regeneration. 
Once the plan satisfies the engineer’s criteria, approval is given to proceed to the implementation phase.

\subsubsection{Coding}
The AI coding agent generates the code and iteratively integrates feedback from validation tools such as compilers and linters until the code meets the required standards (i.e., passes the validation). 
Then, the software engineer is accountable for reviewing the code, making necessary modifications, or providing feedback for further refinement. 
Once the generated code is approved, the software engineer may proceed to submit a pull request.

\subsubsection{Pull Request}
Upon approval, the code changes are assembled into a pull request on Bitbucket for review, adhering to the team's standard workflow. 
Alternatively, a new branch can be created from the generated code to allow for additional modifications.

In our previous publication~\cite{takerngsaksiri2024human}, we conducted an in-depth evaluation of functional correctness and practitioner acceptance using a multi-stage evaluation framework. 
However, aspects of code quality -- particularly code readability, which directly impacts the maintainability of the code~\cite{buse2009learning, dorn2012general, scalabrino2016improving, piantadosi2022evolution} -- remained largely unexamined. 
In this paper, we focus on a detailed assessment of code quality beyond user acceptance (e.g., the merge ratio of pull requests).
\section{Motivation and Related Works}\label{sec_related_works}

In this section, we discuss motivation and related work to formulate the research questions.

% \textit{\textbf{Motivation}}. 
\subsection{Motivation}
Code readability is a cornerstone of software development at Atlassian, particularly as we integrate LLMs into Jira and other platforms~\cite{jiraai2024, takerngsaksiri2024human}.
With the adoption of LLM-powered tools, the size and complexity of codebases are expected to grow rapidly as the majority of the codebase could be generated by LLMs~\cite{adam2004}.
This paradigm shift raises concerns among Atlassian software engineers and customers regarding the readability of LLM-generated code, which may increase technical debts, incur maintenance costs, and negatively affect enterprise coding standards.
Therefore, it is of utmost importance to understand whether readability retains its value in this LLM-assisted paradigm, and if so, why? And, how readable is LLM-generated code compared to human-written code? 

% \textit{\textbf{Related Works}}. 
\subsection{Related Works}
Code readability is a property that influences how easily a given piece of code can be read and understood~\cite{buse2008metric}.
Over the past 20 years~\cite{tashtoush2013impact, coleman2018aesthetics, fakhoury2018effect, mannan2018towards, alawad2019empirical, johnson2019empirical}, prior studies have focused on investigating factors influencing code readability and developing metrics, models, and tools to quantify code readability.
For example, Buse and Weimer~\cite{buse2008metric, buse2009learning} found that the average lines of code and average number of identifiers per line are closely associated with code readability.
% \textbf{Gam, can you highlight the findings, not what they do, but focus on what they found} are associated with code readability.
Similarly, Posnett~\ea~\cite{posnett2011simpler} identified associations between readability and metrics such as the number of lines, Halstead Volume, and character entropy. 
Furthermore, Scalabrino~\ea~\cite{scalabrino2016improving, scalabrino2018comprehensive} emphasized that readability is influenced not only by code structure but also by textual features such as code comments. 
Moreover, Alawad~\ea~\cite{alawad2019empirical} observed a negative correlation between readability and code complexity, reporting that higher complexity is often associated with lower readability. 

In the age of large language models (LLMs), studying code readability is more crucial than ever. 
While LLMs have now been seamlessly integrated into software development tools, workflows, and platforms like GitHub Copilot~\cite{copilot2024}, IntelliJ IDEA~\cite{intellij2025}, ChatGPT~\cite{openai2023gpt4}, and Atlassian's Jira AI~\cite{jiraai2024, takerngsaksiri2024human}, they also introduce new challenges. 
Practitioners often raised concerns about the quality of LLM-generated code.
For example, Liu~\ea~\cite{liu2024refining} found that 47\% of ChatGPT-generated code snippets suffer from maintainability issues.
Majdinasab~\ea~\cite{majdinasab2024assessing} discovered that 27\% of code suggested by GitHub Copilot contains code vulnerabilities.
Similarly, Perry~\ea~\cite{perry2023users} found that users who had access to an AI assistant wrote significantly less secure code than those without access.
Such concerns often lead to a lack of trust in the adoption of LLM-powered software development tools.
% In particular, when LLMs generate code that is not easily readable, developers may be hesitant to adopt these tools into their workflows, fearing that they will introduce more problems than they solve.

While many studies flag issues on LLM-generated code in the security aspect, it is still unclear on the readability aspect of how readable is LLM-generated code compared to human-written code.
Recently, Madi~\cite{al2022readable} found that code generated by GitHub Copilot is comparable in complexity and readability to code written by human pair programmers in the live coding of a controlled environment.
However, little is known about the code readability of LLM-generated code on production in real-world scenarios and how the practitioners perceived code readability in the age of large language models.
The importance of code readability and the gap in literature lead us to formulate the following research questions: 
% \textbf{(RQ1) \emph{\rqone} and (RQ2) \emph{\rqtwo}}
\begin{enumerate}[label=\textbf{RQ\arabic{*}.}, left=0pt]
    \item \textbf{\emph{\rqone}}
    \item \textbf{\emph{\rqtwo}}
\end{enumerate}

\section{(RQ1) \rqone}\label{sec_exp_design}

\subsection{Approach}
The practitioners' survey aims to explore four main topics of code readability: the importance, the challenges, the state of practice and the readability of LLM-generated code.
In this section, we describe the design of our survey study and the participant selection process.

% \textbf{Survey} We design the 15-minute Google Form to gather data and analyze responses by the following method. 
% Excluding the demographic questions, the survey consists of 13 questions (see Table~\ref{tab:survey_questions}).

\begin{table}[th]
    % \centering
    \caption{Survey questions (excluding demographics).}
    \label{tab:survey_questions}
    % \resizebox{\linewidth}{!} { % 
    \begin{tabular}{l|p{0.8\linewidth}}
        \textbf{Item} & \textbf{Question} \\
        \hline
        & \textbf{The Importance of Code Readability} \\
        Q1.1$^\dagger$ & How much of your time is devoted to reading code in your professional practice? \\
        Q1.2$^\ddagger$ & How important is code readability in your opinion? \\
        Q1.3$^\ddagger$ & What motivates you to prioritize code readability improvement in your professional practice? \\
        \hline
        & \textbf{The Challenges of Code Readability} \\
        Q2.1$^\dagger$ & How often do you prioritize code readability in your professional practice? \\
        Q2.2$^\ddagger$ & What factors prevent you from prioritizing the improvement of code readability in your professional practice? \\
        \hline
        & \textbf{The State of Code Readability} \\
        Q3.1$^\ddagger$ & What are the factors you consider when improving the code readability?\\
        Q3.2$^\ddagger$ & Would you consider adopting Generative AI to improve code readability? \\
        Q3.3 & How do you improve your code readability in your professional practice? \\
        Q3.4* & If you improve code readability from automated tools, please kindly specify the name of the tools. \\
        \hline
        & \textbf{The Code Readability of Generative AI} \\
        Q4.1$^\dagger$ & How often do you use Generative AI for code generation/completion in your professional practice? \\
        Q4.2 & To what extent do you think model-generated code is readable compared to human code in professional practice? \\
        % Q4.3 & In Figure 1, which code do you think is more readable? \\
        % Q4.4* & Please justify your answer to Q4.3. \\
        \hline
    \end{tabular}
    % }
    \newline
    \scriptsize{$\dagger$ Likert scale of frequency}
    \scriptsize{$^\ddagger$ Likert scale of agreement 
    % $\bigcirc$ Strongly agree, $\bigcirc$ Somewhat agree, $\bigcirc$ Neutral, $\bigcirc$ Somewhat disagree, $\bigcirc$ Strongly disagree
    }
    \scriptsize{*Open-ended questions.}
\end{table}

\textbf{Survey.} We design the 15-minute Google Form with four main sections: the importance of code readability, the challenges of code readability, the state of code readability, and the readability of LLM-generated code.
The themes for the Likert scales of agreement questions (e.g., motivations for code readability) are derived from the literature review.
Specifically, we categorized factors identified in the literature into themes, which formed the design of our survey questions (i.e., Q1.3, Q2.2, Q3.1).
Excluding the demographic questions, the survey questions are shown in Table~\ref{tab:survey_questions}.
% The full list of our survey questions and themes can be found on our GitHub repository~\cite{coderead2024}.
The lists of derived themes are shown in Table~\ref{tab:lit-important} (Q1.3 and Q2.2) and Table~\ref{tab:lit-factor} (Q3.1).
Lastly, the survey obtained Ethical Permission from the Monash University Human Research Ethics Committee (MUHREC, Project ID 42299) before conducting the research.
% shown in Table~\ref{tab:lit-important} and Table~\ref{tab:lit-factor}.
% Consequently, we design our survey according to these themes.
% The importance (Q1.2) and challenges (Q2.2) are designed using Table~\ref{tab:lit-important}, while the state of code readability (Q3.1) uses Table~\ref{tab:lit-factor}.
% For the Generative AI section, we select a sample code (See Figure~\ref{fig:q43-code}) from the sample ID 20 of the HumanEval-X dataset.
% We used ChatGPT to generate the answer for Generative AI code as the model is the simplest LLM that everyone can access.
% Then, without informing participants which code is the Generative AI code, we ask the participants to select the more readable code (Q4.3) and justify the answer (Q4.4).
% The survey obtained Ethical Permission from the Monash University Human Research Ethics Committee (MUHREC, Project ID 42299) before conducting the research.

\begin{table}[t]
    \centering
    % \caption{Factors Influencing Code Readability Improvement \\Identified in the Literature.}
    \caption{The Importance of Code Readability \\Mentioned in the Literature.}
    \begin{adjustbox}{width=\linewidth}
    {\normalsize   
    \begin{tabular}{c|p{0.4\linewidth}|p{0.4\linewidth}} %{l|l|l}
        \textbf{No.} & \textbf{Description} & \textbf{Source}  \\
        \hline
        1 & Bugs Preventing & \cite{piantadosi2022evolution, scalabrino2018comprehensive} \\
        \hline
        2 & Code Complexity & \cite{fakhoury2019improving, tashtoush2023notional, alawad2019empirical, holst2021importance, fakhoury2018effect} \\
        \hline
        3 & Code Comprehension & \cite{buse2009learning, posnett2011simpler, dorn2012general, scalabrino2018comprehensive, piantadosi2022evolution, sampaio2016software, fakhoury2018effect, johnson2019empirical} \\
        \hline
        4 & Code Maintenance and Evolution & \cite{buse2008metric, buse2009learning, dorn2012general, scalabrino2016improving, scalabrino2018comprehensive, posnett2011simpler, sedano2016code, karanikiotis2020data, piantadosi2022evolution, sampaio2016software, lee2013study, piantadosi2020does, johnson2019empirical} \\
        \hline
        5 & Code Standard \& Aesthetic & \cite{lee2013study, borstler2016beauty, coleman2018aesthetics} \\
        \hline
        6 & Design Quality & \cite{scalabrino2018comprehensive, fakhoury2019improving, mannan2018towards, tashtoush2013impact, piantadosi2022evolution, sampaio2016software} \\
        \hline
        7 & Team Collaboration & \cite{holst2021importance, sampaio2016software, lee2013study, piantadosi2020does, johnson2019empirical} \\
        \hline
    \end{tabular}
    }
    \end{adjustbox}
    \label{tab:lit-important}
\end{table}

\begin{table}[t]
    \centering
    \caption{Coding Factors Related to Code Readability \\ Identified in the Literature.}
    \begin{adjustbox}{width=\linewidth}{\normalsize   
    \begin{tabular}{c|p{0.6\linewidth}|p{0.3\linewidth}} %{l|l|l}
        \textbf{No.} & \textbf{Description} & \textbf{Source}  \\
        \hline
        1 & Code length (e.g., length of the identifier, line, function, file, etc.) & 
        \cite{buse2008metric, buse2009learning, dorn2012general, tashtoush2013impact, posnett2011simpler} \\
        \hline
        2 & Code structure/format (e.g., indentation, use of brackets, blank lines, etc.) & 
        \cite{buse2008metric, buse2009learning, tashtoush2013impact, dorn2012general} \\
        \hline
        3 & Code style/standard (e.g., PEP8, ESLint, Linter, etc.) & \cite{lee2013study, borstler2016beauty, coleman2018aesthetics, fakhoury2019improving} \\
        \hline
        4 & Comment and documentation & \cite{buse2008metric, buse2009learning, dorn2012general, tashtoush2013impact, scalabrino2018comprehensive, scalabrino2016improving, karanikiotis2020data} \\
        \hline
        5 & Descriptive naming variable and function (i.e., lexicon and semantics) & 
        \cite{tashtoush2013impact,dorn2012general, scalabrino2018comprehensive, scalabrino2016improving, fakhoury2018effect, sedano2016code} \\
        \hline
        6 & Function/file complexity & 
        \cite{johnson2019empirical, fakhoury2018effect, tashtoush2023notional, alawad2019empirical, posnett2011simpler, sedano2016code, karanikiotis2020data} \\
        \hline
    \end{tabular}
    }
    \end{adjustbox}
    \label{tab:lit-factor}
\end{table}

% \begin{figure}
%     \centering
%     \includegraphics[width=\linewidth]{assets/q43-code.pdf}
%     \caption{The Given Code in Q3.4 of the Survey.}
%     \label{fig:q43-code}
%     \Description{}{}
% \end{figure}

% \textit{Analysis.}
% We analyze the open-ended responses (Q4.3) by the first author conducting an open coding~\cite{charmaz2014constructing} analysis to extract the theme of each response.
% Then, the first and the third authors perform a card sorting~\cite{spencer2009card} with the predefined themes to systematically categorize the responses into categories. 
% We assessed the inter-rater reliability by calculating Cohen's Kappa coefficient~\cite{mchugh2012interrater}. 
% As a result, the Kappa value is 0.62, reflecting a substantial level of agreement among the evaluators.

\textbf{Participant.}
We recruit participants through advertisements within Atlassian and public channels, including social media platforms like programming-related Facebook groups and personal LinkedIn networks.
The participants can opt-in to win one of three gift cards of \$20 as a token of appreciation.
Finally, during 3 weeks of advertisement in June 2024, we received 118 survey responses with diverse backgrounds. We summarize the attributes of the participants' demographic in Figure~\ref{fig:demo}.

% \begin{figure}
%     \centering
%     \includegraphics[width=\linewidth]{assets/Demograpic_lang.pdf}
%     \caption{The Main Programming Languages of Participants.}
%     \Description[]{}
%     \label{fig:demo-main-lang}
% \end{figure}

% The demographic questions on main programming languages are multiple-choice answers presented in Figure~\ref{fig:demo-main-lang}.

\subsection{Results}
% The survey of 118 practitioners covers four main sections of practitioners' perception: the importance, the challenges, the state of practice, and the Generative AI of code readability.
Table~\ref{tab:suvey-important} to~\ref{tab:survey-gen} show the results of this research question, covering four themes of code readability based on 118 practitioners' responses.

\textbf{\textit{The Importance:}} \textbf{81\% of practitioners agree that code readability is important and the top motivation to improve code readability is to reduce maintenance cost and effort in the long term.}
Table~\ref{tab:suvey-important} shows that 65\% of practitioners answered spending more than half of their time reading code in their professional practice, while 81\% of them were aware of the importance of code readability.
Particularly, the top three motivations to prioritize code readability are to reduce maintenance costs and effort in the long term, to help the debugging and troubleshooting process, and to improve code comprehension for oneself and others.
% \noindent
\begin{table*}[]
    \centering
    \caption{(RQ1) The Importance of Code Readability from Practitioners' Perspective.}
    \begin{tabular}{lp{0.45\textwidth}|p{0.45\textwidth}}
        \hline
        & \textbf{The Importance of Code Readability} & \\
        \hline
        \textbf{Q1.1} & \textbf{How much of your time is devoted to reading code in your professional practice?} & \likertpcttwo[0.34]{27}{50}{28}{12}{1}\\
        \hline
    \end{tabular}
    \newline
    \vspace*{0.01mm}
    \newline
    \hbox{
    \textcolor{blue1}{\rule{7pt}{7pt}} Always (>75\%) %
    \textcolor{blue2}{\rule{7pt}{7pt}} Often (50-75\%) %
    \textcolor{mygray}{\rule{7pt}{7pt}} Sometimes (25-50\%) %
    \textcolor{organge1}{\rule{7pt}{7pt}} Rarely (<25\%) %
    \textcolor{organge2}{\rule{7pt}{7pt}} Never (0\%)
    }
    \newline
    \vspace{0.01mm}
    \newline
    \vspace{0.1mm}
    \begin{tabular}{lp{0.45\textwidth}|p{0.45\textwidth}}
        \hline
        \textbf{Q1.2} & \textbf{How important is code readability in your opinion?} & \likertpct[0.34]{63}{33}{13}{8}{0}\\
        \hline
        \textbf{Q1.3} & \textbf{What motivates you to prioritize code readability \break improvement in your professional practice?} & \\
        \textbf{A.} & Improve code comprehension for oneself and others. & \likertpct[0.34]{55}{45}{9}{9}{0}\\
        \textbf{B.} & Reduce maintenance cost and effort in the long term. & \likertpct[0.34]{73}{33}{6}{5}{1}\\
        \textbf{C.} & Facilitate reusable and scalable development process. & \likertpct[0.34]{51}{39}{17}{9}{2}\\
        \textbf{D.} & Enhance productivity and collaboration in the team. & \likertpct[0.34]{56}{40}{16}{6}{0}\\
        \textbf{E.} & Reflect professionalism, aesthetic and industrial standard. & \likertpct[0.34]{53}{32}{23}{9}{1}\\
        \textbf{F.} & Help debugging and troubleshooting process. & \likertpct[0.34]{70}{33}{8}{7}{0}\\
        \hline
    \end{tabular}
    \hbox{
    \textcolor{green1}{\rule{7pt}{7pt}} Strongly Agree %
    \textcolor{green2}{\rule{7pt}{7pt}} Agree %
    \textcolor{mygray}{\rule{7pt}{7pt}} Neutral %
    \textcolor{red1}{\rule{7pt}{7pt}} Disagree %
    \textcolor{red2}{\rule{7pt}{7pt}} Strongly Disagree
    }
    \label{tab:suvey-important}
    % \noindent\makebox[\linewidth]{\rule{\textwidth}{0.4pt}}
\end{table*}

% \noindent
\begin{table*}[]
    \centering
    \caption{(RQ1) The Challenges of Code Readability from Practitioners' Perspective.}
    \begin{tabular}{lp{0.45\textwidth}|p{0.45\textwidth}}
        \hline
        & \textbf{The Challenges of Code Readability} & \\
        \hline
        \textbf{Q2.1} & \textbf{How often do you prioritize code readability in your professional practice?} & \likertpcttwo[0.34]{38}{45}{30}{5}{0}\\
        \hline
    \end{tabular}
    \newline
    \vspace*{0.01mm}
    \newline
    \hbox{
    \textcolor{blue1}{\rule{7pt}{7pt}} Always (>75\%) %
    \textcolor{blue2}{\rule{7pt}{7pt}} Often (50-75\%) %
    \textcolor{mygray}{\rule{7pt}{7pt}} Sometimes (25-50\%) %
    \textcolor{organge1}{\rule{7pt}{7pt}} Rarely (< 25\%) %
    \textcolor{organge2}{\rule{7pt}{7pt}} Never (0\%)
    }
    \newline
    \vspace*{0.5mm}
    \newline
    \vspace{0.5mm}
    \begin{tabular}{lp{0.45\textwidth}|p{0.45\textwidth}}
        \hline
        \textbf{Q2.2} & \textbf{What factors prevent you from prioritizing the improvement of code readability in your professional practice?} & \\
        \textbf{A.} & Time constraints due to project deadlines. & \likertpct[0.34]{40}{53}{14}{9}{2}\\
        \textbf{B.} & Focus on code functionality over code readability. & \likertpct[0.34]{38}{39}{22}{16}{3}\\
        \textbf{C.} & Focus on other performance metrics (e.g., speed and efficiency). & \likertpct[0.34]{28}{41}{29}{17}{3}\\
        \textbf{D.} & Complexity of the codebase or legacy code makes the code readability improvement challenging. & \likertpct[0.34]{35}{48}{15}{15}{5}\\
        \textbf{E.} & Resistance from team members or stakeholders. & \likertpct[0.34]{30}{42}{20}{20}{6}\\
        \textbf{F.} & Limited resource or support on standards or best practices on code readability improvement. & \likertpct[0.34]{36}{38}{20}{23}{1}\\
        \hline
    \end{tabular}
    \hbox{
    \textcolor{green1}{\rule{7pt}{7pt}} Strongly Agree %
    \textcolor{green2}{\rule{7pt}{7pt}} Agree %
    \textcolor{mygray}{\rule{7pt}{7pt}} Neutral %
    \textcolor{red1}{\rule{7pt}{7pt}} Disagree %
    \textcolor{red2}{\rule{7pt}{7pt}} Strongly Disagree
    }
    \label{tab:survey-challenge}
    % \noindent\makebox[\linewidth]{\rule{\textwidth}{0.4pt}}
\end{table*}

\textbf{\textit{The Challenges:}} \textbf{79\% of practitioners agree that the top challenging factor preventing them from prioritizing code readability is time constraints due to project deadlines.}
Table~\ref{tab:survey-challenge} shows that 70\% of practitioners answered they always or often prioritize code readability in their professional practice.
However, the top three factors that prevent them from prioritizing the improvement of code readability are the time constraints of project deadlines, the complexity of the codebase or legacy code, and the focus on code functionality over code readability.

\begin{figure}[t]
    \centering
    \includegraphics[width=\columnwidth]{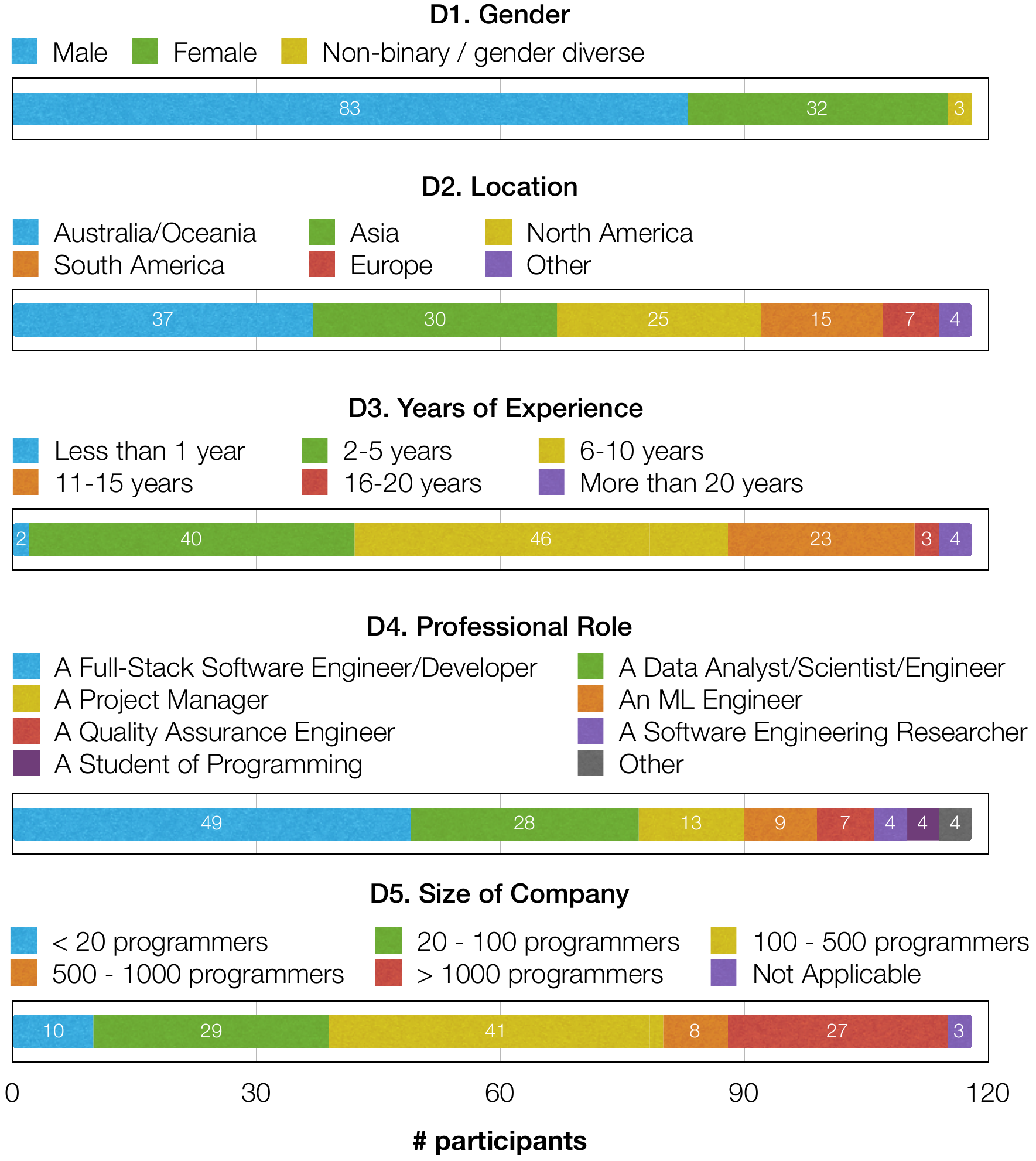}
    \caption{The Demographic of Participants in RQ1 ($n$=118).}
    \label{fig:demo}
\end{figure}

% \noindent
\begin{table*}[]
    \centering
    \caption{(RQ1) The State of Code Readability from Practitioners' Perspective.}
    \begin{tabular}{lp{0.45\textwidth}|p{0.45\textwidth}}
        \hline
        & \textbf{The State of Code Readability} &\\
        \hline
        \textbf{Q3.1} & \textbf{What are the factors you consider when improving the code readability?} & \\
        \textbf{A.} & Code length (e.g., length of identifier, line, function, file, etc.) & \likertpct[0.34]{44}{40}{20}{12}{1}\\
        \textbf{B.} & Code structure/format (e.g., indentation, use of brackets, etc.) & \likertpct[0.34]{56}{35}{19}{7}{1}\\
        \textbf{C.} & Code style/standard (e.g., PEP8, ESLint, Linter, etc.) & \likertpct[0.34]{46}{41}{18}{10}{3}\\
        \textbf{D.} & Comment and documentation & \likertpct[0.34]{46}{53}{14}{5}{0}\\
        \textbf{E.} & Descriptive naming variable/function & \likertpct[0.34]{58}{44}{10}{5}{0}\\
        \textbf{F.} & Function/file complexity & \likertpct[0.34]{59}{38}{15}{6}{0}\\
        \hline
        \textbf{Q3.2} & \textbf{Would you consider adopting LLMs to improve code readability?} & \likertpct[0.34]{47}{38}{23}{8}{2}\\
        \hline
    \end{tabular}
    \newline
    \vspace{0.01mm}
    \newline
    \hbox{
    \textcolor{green1}{\rule{7pt}{7pt}} Strongly Agree %
    \textcolor{green2}{\rule{7pt}{7pt}} Agree %
    \textcolor{mygray}{\rule{7pt}{7pt}} Neutral %
    \textcolor{red1}{\rule{7pt}{7pt}} Disagree %
    \textcolor{red2}{\rule{7pt}{7pt}} Strongly Disagree
    }
    \newline
    \vspace*{0.01mm}
    \newline
    \begin{tabular}{lp{0.45\textwidth}|p{0.45\textwidth}}
        \hline
        \textbf{Q3.3} & \textbf{How do you improve your code readability in your professional practice?} & \likertpcttwo[0.34]{48}{32}{4}{14}{20}\\
        \hline
    \end{tabular}
    \newline
    \vspace*{0.01mm}
    \newline
    \hbox{
    \textcolor{blue1}{\rule{7pt}{7pt}} Manually by peer review comments %
    \textcolor{blue2}{\rule{7pt}{7pt}} Manually self-improve code %
    \textcolor{mygray}{\rule{7pt}{7pt}} Mixed %
    \textcolor{organge1}{\rule{7pt}{7pt}} Static tools %
    \textcolor{organge2}{\rule{7pt}{7pt}} AI-based tools
    }
    \vspace{1.2mm}
    \begin{tabular}{lp{0.46\textwidth}|p{0.44\textwidth}}
        \hline
        % \textbf{Q3.3} & \textbf{How do you improve your code readability in your professional practice?} \\
        % \textbf{A.} & Manually by peer review comments (e.g., code review) & \barplt{48}\\
        % \textbf{B.} & Manually self-improve code & \barplt{32} \\
        % \textbf{C.} & AI-based tools & \barplt{20}\\
        % \textbf{D.} & Static tools (e.g., AutoPEP8) & 
        % \barplt{14}\\
        % \textbf{E.} & Both manually and AI-based tools & 
        % \barplt{1}\\
        % \textbf{F.} & All of the above & 
        % \barplt{3}\\
        % \hline
        \textbf{Q3.4*} & \textbf{If you improve code readability from automated tools, please kindly specify the name of the tools.} \\
        \textbf{A.} & ChatGPT & \barplt{12}\\
        \textbf{B.} & ESLint & \barplt{5} \\
        \textbf{C.} & Prettier & \barplt{5}\\
        \textbf{D.} & Copilot & \barplt{3}\\
        \textbf{E.} & KtLint & \barplt{3}\\
        \textbf{F.} & Black & \barplt{2}\\
        \hline
    \end{tabular}
    \label{tab:survey-state}
    \footnotesize{*Open-ended questions.}
    % \noindent\makebox[\linewidth]{\rule{\textwidth}{0.4pt}}
\end{table*}

\textbf{\textit{The State of Practice:}} \textbf{While manually improving code readability is still a dominant method with 68\% of practitioners selection, 72\% of practitioners agree to consider adopting LLMs to improve code readability.}
Table~\ref{tab:survey-state} shows that the most popular method for improving code readability in professional practice is still done manually by either peer review (41\%) or self-improvement (27\%).
However, 17\% of the practitioners have started to adopt LLMs to improve the code readability.
The most commonly indicated tool is ChatGPT (n=12).
Although the majority of the practitioners currently improve code manually, 72\% of them are open to adopting LLMs to assist with code readability improvement.
Additionally, the top three factors that practitioners consider when improving code readability are the descriptive naming variables and functions, the comments and documentation, and the complexity of functions or files.

% \noindent
\begin{table*}[]
    \centering
    \caption{(RQ1) The Readability of LLM-generated Code from Practitioners' Perspective.}
    \begin{tabular}{lp{0.45\textwidth}|p{0.45\textwidth}}
        \hline
        & \textbf{The Readability of LLM-generated Code} & \\
        \hline
        \textbf{Q4.1} & \textbf{How often do you use LLMs for code generation/completion in your professional practice?} & \likertpcttwo[0.34]{20}{44}{25}{23}{6}\\
        \hline
    \end{tabular}
    \newline
    \vspace{0.5mm}
    \newline
    \hbox{
    \textcolor{blue1}{\rule{7pt}{7pt}} Always (>75\%) %
    \textcolor{blue2}{\rule{7pt}{7pt}} Often (50-75\%) %
    \textcolor{mygray}{\rule{7pt}{7pt}} Sometimes (25-50\%) %
    \textcolor{organge1}{\rule{7pt}{7pt}} Rarely (<25\%) %
    \textcolor{organge2}{\rule{7pt}{7pt}} Never (0\%)
    }
    \newline
    \vspace*{0.5mm}
    \newline
    \begin{tabular}{lp{0.45\textwidth}|p{0.45\textwidth}}
        \hline
        \textbf{Q4.2} & \textbf{To what extent do you think LLM-generated code is readable compared to human code in professional practice?} & \likertpcttwo[0.34]{32}{0}{40}{0}{46}\\
        \hline
    \end{tabular}
    \newline
    \vspace*{0.01mm}
    \newline
    \hbox{
    \textcolor{blue1}{\rule{7pt}{7pt}} Human-written code is more readable %
    % \textcolor{blue2}{\rule{7pt}{7pt}} Manually self-improve code %
    \textcolor{mygray}{\rule{7pt}{7pt}} Both are similar %
    % \textcolor{organge1}{\rule{7pt}{7pt}} Static tools %
    \textcolor{organge2}{\rule{7pt}{7pt}} LLM-generated code is more readable
    }
    % \newline
    % \vspace*{1.5mm}
    % \newline
    % \begin{tabular}{lp{0.46\textwidth}|p{0.435\textwidth}}
    %     % \hline
    %     % \textbf{Q4.2} & \textbf{To what extent do you think model-generated code is readable compared to human code in professional practice?} \\
    %     % \textbf{A.} & Both are similarly readable & \barplt{39}\\
    %     % \textbf{B.} & Model-generated code is more readable & \barplt{46} \\
    %     % \textbf{C.} & Human written code is more readable & \barplt{32}\\
    %     % \textbf{D.} & None is readable & 
    %     % \barplt{1}\\
    %     \hline
    %     % \textbf{Q4.3} & \textbf{In Figure~\ref{fig:q43-code}, which code do you think is more readable?} \\
    %     % \textbf{A.} & Both are similarly readable & \barplt{25}\\
    %     % \textbf{B.} & Code A is more readable than code B & \barplt{70} \\
    %     % \textbf{C.} & Code B is more readable than code A & \barplt{17}\\
    %     % \textbf{D.} & None is readable & 
    %     % \barplt{6}\\
    %     % \hline
    %     \textbf{Q4.3*} & \textbf{Please justify your answer to Q4.2.} \\
    %     \textbf{A.} & Logic complexity & \barplt{25}\\
    %     \textbf{B.} & Code Length & \barplt{15} \\
    %     \textbf{C.} & A number of loop condition & \barplt{12}\\
    %     \textbf{D.} & Code structure/format & 
    %     \barplt{12}\\
    %     \textbf{E.} & Descriptive naming variable & 
    %     \barplt{8}\\
    %     \textbf{F.} & Code comment & 
    %     \barplt{2}\\
    %     \hline
    % \end{tabular}
    % \newline
    % \footnotesize{*Open-ended questions.}
    \label{tab:survey-gen}
    % \noindent\makebox[\linewidth]{\rule{\textwidth}{0.4pt}}
\end{table*}

\textbf{\textit{The Readability of LLM-generated Code:}} 
\textbf{39\% of practitioners answer they perceive the LLM-generated code as more readable than the human-written code.}
% \textbf{By showing a code pair of anonymous origin, 59\% of practitioners perceived that the ChatGPT code is more readable than the human solutions.}
Table~\ref{tab:survey-gen} shows that 54\% of practitioners answered using LLMs to generate or complete code more than half of their time in professional practice, indicating the current trend of high usage in LLM-generated code.
Meanwhile, 39\% of practitioners answered they perceive the LLM-generated code as more readable than the human-written code, followed by 34\% of them answered both are similarly readable.
% In fact, by showing a code pair of unidentified origin, 59\% of practitioners select the ChatGPT-generated code as more readable than the human-written code.
The top justifications from the follow-up open-text question are logic complexity (n=25), code length (n=15), a number of loop conditions (n=12) and code structure/format (n=12).
Specifically, a practitioner who voted LLM-generated code as more readable indicated that \textit{``Human code has too many logic conditions to check''}.
On the other hand, a practitioner who voted Human code as more readable indicates that \textit{``I find it easier to follow along.''}
Interestingly, some practitioners reason that \textit{``Actually, Human code is easier to understand in terms of solution, but having a lot of nested loop and conditions makes it harder to read and debug.''}, indicates that understanding code and reading code can be as well on the different scales of decision.
% Specifically, a practitioner who voted Generative AI code as more readable indicated that \textit{``Nested for loop make it more complicated to understand''}, while another said \textit{``Human code has too many logic conditions to check''}.
% On the other hand, a practitioner who voted Human code as more readable indicates that \textit{``I find it easier to follow along with exactly what each variable represents and how the lists are being compared.''}
% Interestingly, some practitioners reason that \textit{``Actually, Human code is easier to understand in terms of solution, but having a lot of nested loop and conditions makes it harder to read and debug.''}, indicates that understanding code and reading code can be as well on the different scales of decision.

% \begin{figure}
%     \centering
%     \includegraphics[width=\linewidth]{assets/q43-human-vs-ai.pdf}
%     \caption{Practitioners' Perspective on more readable code from examples of Figure~\ref{fig:q43-code}.}
%     \Description[]{}
%     \label{fig:q43-stat}
% \end{figure}

% \begin{figure}
%     \centering
%     \includegraphics[width=0.9\linewidth]{assets/q4-open-text.pdf}
%     \caption{Reasons of Readable Code by Practitioners for Q4.3.}
%     \Description[]{}
%     \label{fig:q4-reason}
% \end{figure}

% \begin{figure*}
%     \centering
%     \includegraphics[width=0.95\textwidth]{assets/Code Readability Likert-1.pdf}
%     \caption{Caption}
%     \label{fig:likert-all}
% \end{figure*}

\section{(RQ2) \rqtwo}\label{sec_exp_result}

% \subsection{Motivation}
% Previous work~\cite{al2022readable} conducted an experiment to evaluate the readability of model-generated code via static analysis and eye tracing while live coding.
% However, the study focuses mainly on the specific context in controlled environments.
% Our case study aims to examine the code readability in the real-world production of Atlassian.

\subsection{Approach}
The empirical study aims to understand code readability via static code analysis.
% We mainly select the evaluation metrics following the existing study on code readability~\cite{al2022readable}.
In this section, we describe the design of our case study on HULA~\cite{takerngsaksiri2024human} over our internal dataset.

\begin{figure}[t]
    \centering
    \includegraphics[width=\linewidth]{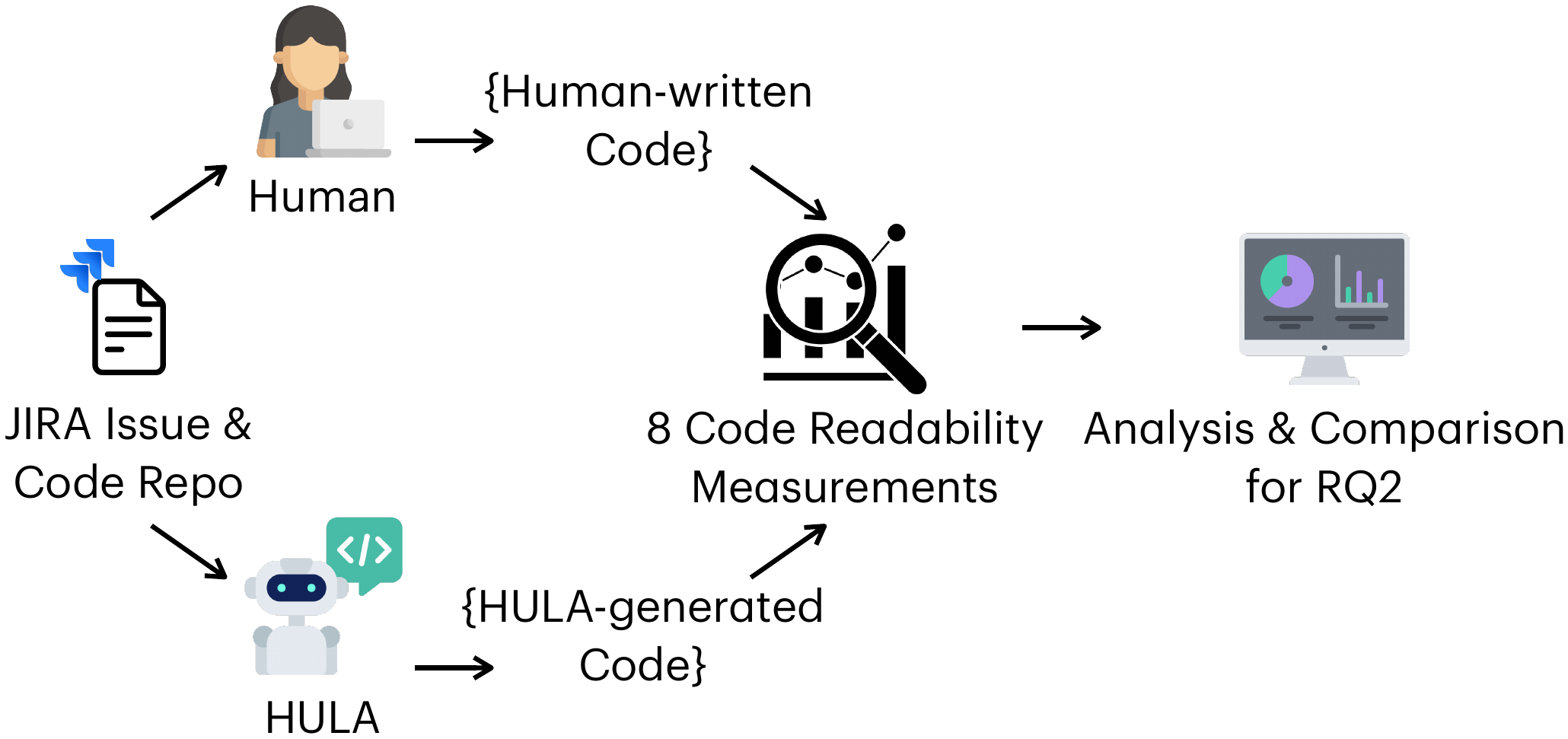}
    \caption{The Overview of the RQ2 Empirical Study Comparing Human-written Code to HULA-generated Code on Code Readability Measures.}
    \label{fig:rq2-overview}
\end{figure}

\textbf{Overview.}
% \textcolor{blue}{to be added}
Figure~\ref{fig:rq2-overview} shows the overview of the approach in this research question.
Given a JIRA issue and a corresponding code repository, a software engineer writes a code, while AI coding agents, HULA, generate the codes to resolve the JIRA issue.
Then, the human-written code and the HULA-generated code are assessed via eight code readability measurements.
Lastly, we compare the results using the statistical analysis of the significance of Mann-Whitney U Test~\cite{mann1947test} and the effect size of Cliff's Delta~\cite{cliff1993dominance}.

\textbf{Large Language Model.}
We evaluate the code readability using our Human-in-the-loop Software Development Agents framework, \textit{HULA}~\cite{takerngsaksiri2024human}.
{HULA} is Atlassian's LLM-based agents framework for software development available internally in Jira Software.
Specifically, the framework is used to generate a code patch given a code repository and a description of the Jira issue to be resolved.
% The framework is used on the internal dataset to generate coding files to solve Jira Issues. 
In this version, HULA uses GPT-4~\cite{openai2023gpt4} as a based LLM and the framework is evaluated without additional human feedback.

\textbf{Datasets.}
We evaluate HULA via the \textit{internal dataset}~\cite{takerngsaksiri2024human}, which includes a set of completed software development tasks (i.e., Jira issues, repositories and corresponding pull requests from BitBucket) at Atlassian.
The dataset consists of 144 Jira issues with 250 coding files spanning six programming languages: TypeScript, Java, Kotlin, Python, Go and Scala.
The distribution of programming languages is illustrated in Figure~\ref{fig:data-lang-portion}.

\begin{figure}[h]
    \centering
    \includegraphics[width=0.6\linewidth]{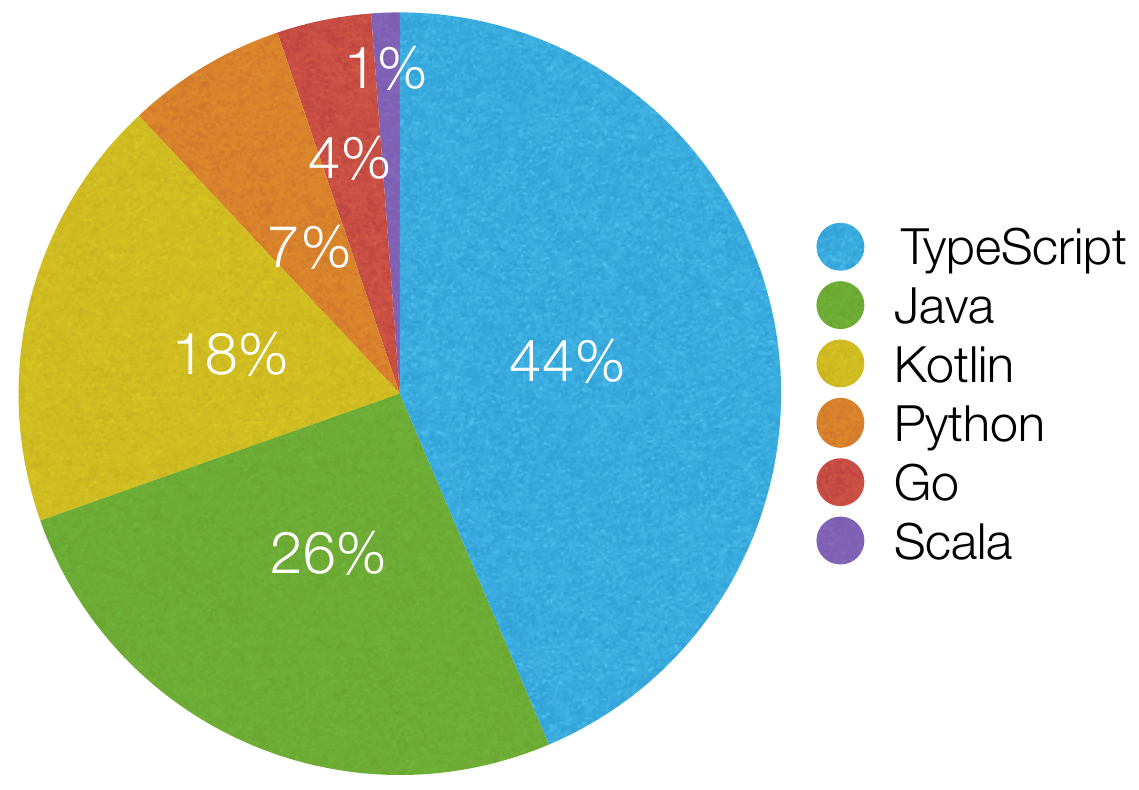}
    \caption{The Distribution of Programming Languages in the Dataset (n=250).}
    \label{fig:data-lang-portion}
\end{figure}

% based on the most popular programming languages online\footnote{https://spectrum.ieee.org/top-programming-languages-2024} and on our codebase.

% \begin{itemize}
%     % \item \textbf{HumanEval-X dataset}~\cite{zheng2023codegeex} is an open-source benchmark for assessing the multilingual ability of code generation models.
%     % The dataset is extended from the HumanEval dataset~\cite{chen2021codex} which is a set of 164 hand-written programming problems in Python.
%     % HumanEval-X broadens the scope by including five programming languages: Python, Java, JavaScript, C++, and Go, comprising a total of 820 (5*164) crafted data samples.
%     % In this study, we use the standard prompt structure of HumanEval on all the LLMs.
%     % Only the coding part is extracted, while the explanation part is discarded if presented.
    
%     \item \textbf{Internal dataset}~\cite{takerngsaksiri2024human} is a set of completed software development tasks (i.e., Jira issues) at Atlassian.
%     % Originally, the dataset is used to assess the software development agents' ability to resolve development issues.
%     This internal dataset consists of 144 Jira issues with 250 coding files of corresponding pull requests (PRs) from BitBucket.
%     The coding files include six programming languages: TypeScript, Java, Kotlin, Python, Go and Scala.
%     % The proportion of the programming languages is shown in Figure~\ref{fig:data-lang-portion}.
% \end{itemize}

\textbf{Code Readability Measurement.}
% Selecting metrics to assess code readability is challenging; many studies~\cite{buse2008metric, scalabrino2018comprehensive, dorn2012general} attempt to propose code readability metrics with various formulas and models.
% However, most of the metrics are available in certain programming languages such as Java.
% we cautiously select metrics that can be used in multilingual, expanding the context to cover eight programming languages, including both an open-source and a proprietary dataset.
Following the recent study on code readability~\cite{al2022readable}, we cautiously select the following eight static analysis metrics that are language-agnostic (i.e., can be used in multilingual) via the Multimetric library\cite{multimetric2024}.

First, we use \textit{Line of Code} to evaluate the total number of lines of code in the file, expressing the file size.
The \textit{Comment Ratio} is used to evaluate the percentage of the lines of comment to code, expressing the descriptive of the code.
Then, \textit{Cyclomatic Complexity}~\cite{mccabe1976complexity} is used to evaluate the number of decision paths of the code, expressing the code complexity. 
Next, \textit{Maintainability Index}~\cite{Oman1992maintainability} is used to measure how maintainable the code is.
Lastly, we utilize \textit{Halstead Metrics}~\cite{halstead1977elements}, a set of metrics designed to quantify various aspects of code. These metrics are employed to evaluate key attributes, including \textit{Difficulty} (the difficulty of the program to write or understand), \textit{Vocabulary} (the number of unique operators and unique operands occurrence), \textit{Volume} (the program length and vocabulary size), and \textit{Time Required} (the estimated time for implementation).

Below, we list the detailed calculations of each metric.

% \begin{itemize}[leftmargin=*]
%     \item[] \textit{Line of Code} is the total number of lines of code.
%     \item[] \textit{Comment Ratio} is the percentage of the lines of comment to the lines of code.
%     \item[] \textit{Cyclomatic Complexity}~\cite{mccabe1976complexity} is the number of decisions path of code according. 
%     \item[] \textit{Maintainability Index}~\cite{Oman1992maintainability} is a metric to measure how maintainable the code is.
%     \item[] \textit{Halstead Metrics}~\cite{halstead1977elements} is a set of metrics to evaluate code quality, i.e., Difficulty, Vocabulary, Volume, and Time Required.
% \end{itemize}

\begin{figure*}
    \centering
    \includegraphics[width=0.9\textwidth]{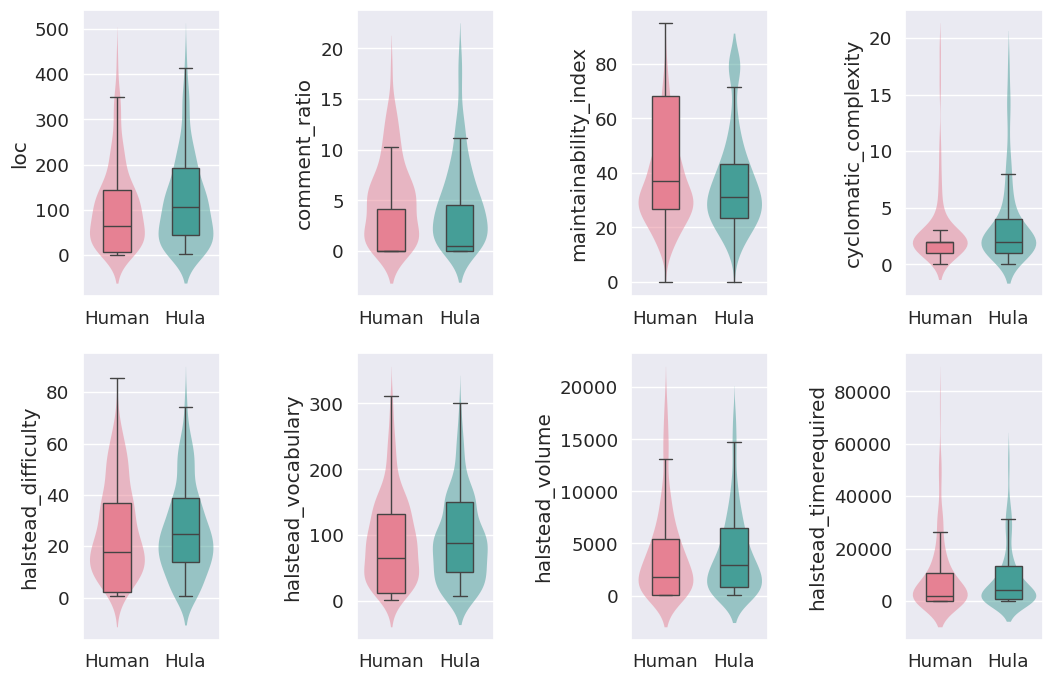}
    \caption{(RQ2) The Scores of Code Readability Measures between HULA-generated and Human-written Code.}
    % \Description[Code Readability Measurement]{The Scores of Code Readability Measurement between Human and HULA Code on the Internal Dataset.}
    \label{fig:boxplot-hula}
\end{figure*}

\begin{figure*}
    \centering
    \includegraphics[width=\textwidth]{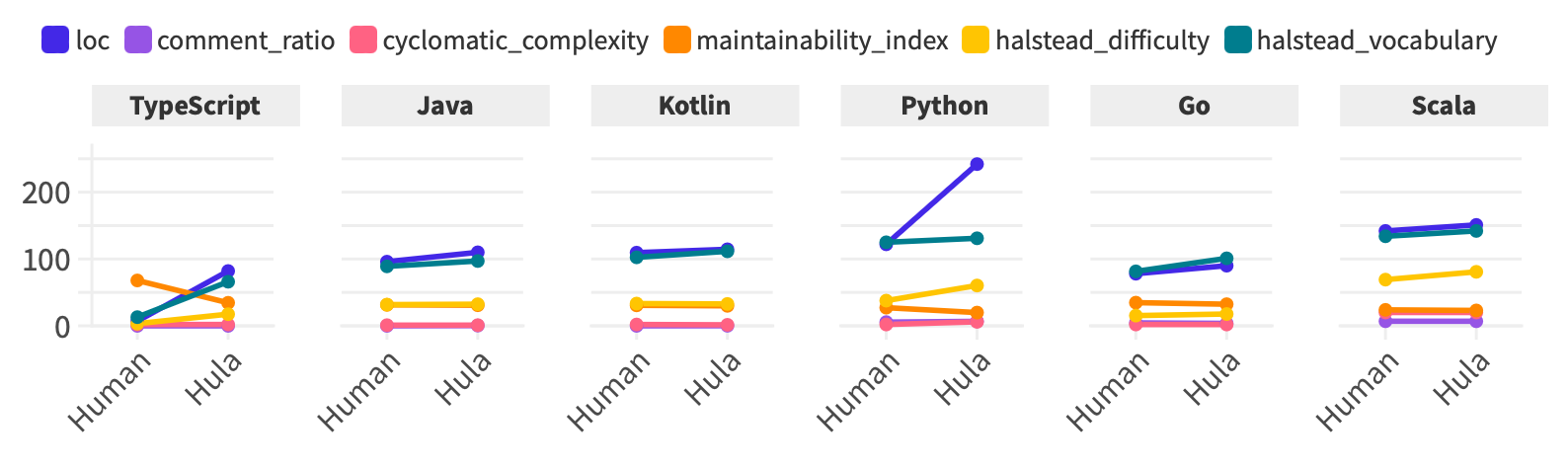}
    % \Description[Median Score of Internal by Language]{The Median Scores of Human and Generative AI code on the Atlassian Internal Dataset by Programming Languages.}
    \caption{(RQ2) The Median Scores of Code Readability Measures between Human-written and HULA-generated code by Programming Languages.}
    \label{fig:hula-by-lang}
\end{figure*}

\begin{enumerate}
    \item \textbf{Line of Code} is the total number of lines of code.
    \item \textbf{Comment Ratio} is the percentage of the lines of comment to the lines of code.
    \item \textbf{Cyclomatic Complexity}~\cite{mccabe1976complexity} or McCabe number is the number of decisions path of code according. Specifically, the complexity is defined as $C = E - N + 2$, where $E$ is the number of edges on the control flow graph and $N$ is the number of nodes on the control flow graph.
    \item \textbf{Maintainability Index}~\cite{Oman1992maintainability} is a metric to measure how maintainable the code is. We use the Microsoft formula~\cite{microsoft2022maintainability} defined as $Max(0,(171 - 5.2 * ln(Halstead Volume) - 0.23 * (McCabe) - 16.2 * ln(LoC))*100 / 171)$.
    \item \textbf{Halstead Difficulty}~\cite{halstead1977elements} defines as $D = (n1 / 2) * (N2 / n2)$, where $n1$ is the number of distinct operators, $n2$ is the number of distinct operands, $N1$ is the total number of operators, and $N2$ is the total number of operands.
    \item \textbf{Halstead Vocabulary}~\cite{halstead1977elements} defines as $n = n1 + n2$.
    \item \textbf{Halstead Volume}~\cite{halstead1977elements} defines as $V = (N1+N2) * log_2(n)$.
    \item \textbf{Halstead Time Required}~\cite{halstead1977elements} defines as $T = (D*V) / 18$
\end{enumerate}

% \subsubsection{Analysis}
% \begin{itemize}
%     \item \textbf{Mann-Whitney Test}
%     \item \textbf{Cliff’s Delta}
% \end{itemize}

\subsection{Results}

Figure~\ref{fig:boxplot-hula} shows the results of this research question, covering the case study of HULA on our internal dataset.
% The analysis is done on two datasets: HumanEval-X, which represents the open-source dataset, and Atlassian internal, which represents the real-world production code.
% Figure~\ref{fig:boxplot-humaneval} and~\ref{fig:boxplot-hula} show the results of this research question.
Table~\ref{tab:rq2-significant} shows the statistical significance and the effect size of the result.
Lastly, Figure~\ref{fig:hula-by-lang} shows the detailed analysis of code readability measurement by programming languages.

% \textbf{On HumanEval-X, the Generative AI significantly generate longer code including more lines of code and comment ratio, leading to a significant decline in the maintainability index.} 
% Figure~\ref{fig:boxplot-humaneval} shows that the code generated from LLMs is always significantly longer (i.e., higher lines of code and comment ratio).
% Specifically, GPT-3.5 exhibits the most similar trend to humans with a small effect size of difference, while GPT-4 is the most extreme having the highest code length, vocabulary, volume, time required and the lowest maintainability index.
% Nonetheless, the metrics indicating code complexity (i.e., Cyclomatic Complexity and Halstead difficulty) show no statistically significant difference on all LLMs, suggesting an overall comparable complexity to the human-written code. 
% Therefore, these results indicate that the code readability of the Generative AI is mostly affected by the code length more than the complexity of the code itself.

% Additionally, Figure~\ref{fig:gpt4o-by-lang} shows the static analysis results by programming languages of code generated by GPT-4o and written by humans.
% We observe that GPT-4o, the latest model from OpenAI, exhibits a consistent pattern across all the studied programming languages. 
% Specifically, the generated code is significantly longer with an approximately 10\% less maintainable, at least 30\% higher comment ratio, and generally higher values in Halstead's metrics compared to the human-written code.

\textbf{The HULA-generated code is comparable to the human-written code with a negligible to small statistical difference across our eight evaluation metrics.}
Figure~\ref{fig:boxplot-hula} shows that HULA, with a GPT-4 based model, generates code that is highly similar to humans with only a slightly longer code but with no statistically significant difference in Cyclomatic Complexity and comment ratio.
On Halstead's metrics, our LLM-generated code has marginally higher values on difficulty, vocabulary, volume, and time required.
These results indicate that the HULA framework can produce code that closely resembles human-written code in terms of code readability, indicating the promising performance of LLM in production.

\textbf{The HULA-generated code is comparable to the human-written code specifically in Java, Kotlin, Go and Scala programming languages.}
Figure~\ref{fig:hula-by-lang} shows the static analysis results by programming languages of code generated by Hula and written by humans.
The case study on the Hula framework reveals two different patterns across the six studied programming languages. 
First, in Java, Kotlin, Go and Scala programming languages, HULA-generated code shows comparable results across all code readability measures to that written by humans.
Second, in TypeScript and Python programming languages, the Hula-generated code tends to have longer lines of code and slightly lower maintenance scores than the human-written code by small effect sizes.

\begin{table}[t]
\centering
\caption{(RQ2) The mean significance from Mann-Whitney U Test~\cite{mann1947test} and Effect Size measured by the Cliff's Delta~\cite{cliff1993dominance} based on the comparison of HULA and Human code.}
% \begin{adjustbox}{width=\linewidth}
{\normalsize   
\begin{tabular}{l|l|l}

% \textbf{Dataset}   & \multicolumn{2}{c}{\textbf{Internal (n=250)}}\\
% \hline
% \textbf{Approach}  & \multicolumn{2}{c}{\textbf{Hula}}          \\
\hline
\textbf{Metric}    & \textbf{P-Value} & \textbf{Effect Size} \\
\hline
\textbf{Line of Code} & \textless{}0.001 & 0.230 (small)\\
\textbf{Comment Ratio}  & 0.005 &  0.124 (negligible) \\
\textbf{Cyclomatic Complexity}  & 0.622 &  0.015 (negligible) \\
\textbf{Maintainability Index}& \textless{}0.001 & -0.225 (small) \\
\textbf{Halstead Difficulty} & \textless{}0.001 &  0.201 (small)\\
\textbf{Halstead Vocabulary} & \textless{}0.001 &  0.202 (small) \\
\textbf{Halstead Volume}     & \textless{}0.001 &  0.208 (small) \\
\textbf{Halstead Time}       & \textless{}0.001 &  0.205 (small)\\
\hline
\end{tabular}
}
% \end{adjustbox}
\newline
% \footnotesize{$\dagger$ indicates the alternative hypothesis (H1) of Generative AI code has a score significantly less than human code, otherwise significantly more.}
\label{tab:rq2-significant}
\end{table}

\section{Threat to Validity}~\label{sec:threat}
In this section, we disclose the threats to the validity.

\textbf{\textit{Threats to internal validity.}}
We use an online survey to study practitioners' perceptions of code readability.
We design the questions based on literature reviews to mitigate the hallucination on factors related to code readability.
However, we acknowledge the limitations that the justifications might be beyond the literature review which we sourced from.

\textbf{\textit{Threats to external validity.}}
Our survey was derived from the perceptions of 118 practitioners with different backgrounds and work experiences in coding (see Figure~\ref{fig:demo}). 
% The demographic (see Figure~\ref{fig:demo}) shows a great diversity in all attributes.
Nonetheless, we acknowledge that the results may differ when surveying different population groups.
Additionally, the empirical study is a case study of our HULA framework, we acknowledge that the results may differ when experimenting with different models and datasets.

Additionally, the empirical study was conducted on code generation tasks using GPT-4 as a based model and the Atlassian internal dataset. 
Efforts were made to ensure diversity in programming languages and task contexts. 
However, it is acknowledged that the findings may vary when applied to different datasets. 
Therefore, future research is encouraged to investigate code readability across a broader range of contexts.

Lastly, while the HULA framework typically operates with a human-in-the-loop—where software engineers provide feedback to guide the outputs of the LLM-based agent, in this study, all human feedback and intervention were intentionally excluded. 
The effort was made to avoid introducing additional information to the LLM-based agents and to ensure a fair comparison between code written by humans and code generated by LLM-based agents.
% As such, the code generated in this setting may not fully reflect the readability of outputs from fully autonomous agents. 
% This human involvement can improve the final code beyond what the model alone might produce. 
Future studies could compare results from both autonomous and human-in-the-loop settings to better understand their respective trade-offs.

% % Additionally, the empirical study was conducted on code generation tasks over four LLMs (i.e., GPT-3.5, GPT-4, GPT-4o) and two datasets(HumanEval-X and the internal dataset).
% To the best of our effort, we expand the study to cover diverse programming languages and contexts.
% Nonetheless, we acknowledge that the results may differ when conducting the empirical study on different datasets.
% Thus, future research is encouraged to explore the different contexts of code readability.

\textbf{\textit{Threat to construct validity.}}
We use an empirical study with static code analysis as a proxy for code readability measurement between LLM-generated and Human code.
We cautiously follow the existing work on selecting the metrics representing code readability.
However, we acknowledge the limitations that the metrics might not completely represent the code readability from all perspectives and that some inter-dependencies may exist among the metrics.

Last but not least, although we may not be able to disclose the source code of the evaluation dataset due to industrial confidentiality reasons, we are confident that the evaluation offers valuable insight into the practical use of LLMs in real-world development workflows and reflects industrial standards for production-level code, paving the way forward the future of AI-powered software development tools. 

% External validity involves the generalizations of findings made by researchers in hope that their research will be relevant for other populations, settings, etc. (Cozby 2014). Internal validity is instead focused on the structure of a study and the accuracy of the conclusions drawn based on a cause and effect relationship (Andrade 2018; Cozby 2014). Construct validity answers questions about the measurement of a concept or construct (Cozby 2014). It focuses on identifying if the selected concepts or constructs are suitable for the selected variables (Cozby 2014).
\section{Conclusion and Implications}\label{sec_conclusion}

In this paper, we investigate the practitioners' perceptions on the importance, the challenges, and the state of practice of code readability, and investigate the readability of LLM-generated code and human-written code in the context of enterprise software development tasks.
Our findings underscore that readability remains a critical aspect of software development, even in the age of large language models (LLMs). 
Moreover, the code generated by our LLM-powered framework, HULA, is shown to be comparable in readability to human-written code. 
This supports the establishment of appropriate trust and drives the broad adoption of our LLM-powered software development platform.
Practitioners can confidently integrate LLMs into their workflows, knowing that code readability will not be compromised. 
This fosters more effective teamwork, simplifies maintenance tasks, and ensures the long-term success of software projects.

\section{Disclaimer}\label{sec:disclaimer}
The perspectives and conclusions presented in this paper are solely the authors' and should not be interpreted as representing the official policies or endorsements of Atlassian or any of its subsidiaries and affiliates. 
Additionally, the outcomes of this paper are independent of, and should not be construed as an assessment of, the quality of products offered by Atlassian.

\bibliographystyle{IEEEtran}
\bibliography{mybibfile}

\end{document}